\documentclass[superscriptaddress,notitlepage,tightenlines,aps,pra,reprint,twocolumn,amsmath,amssymb,citeautoscript,nofootinbib]{revtex4-1}
\usepackage{graphicx}
\usepackage{dcolumn}
\usepackage{bm}
\usepackage{latexsym,epsfig}
\usepackage{graphicx}
\usepackage{verbatim}
\usepackage{comment}
\usepackage{amsmath}
\usepackage{amssymb}
\usepackage{stmaryrd}
\usepackage{color}
\usepackage{epstopdf}
\usepackage{grffile}
\usepackage{mathtools}
\usepackage{physics}
\DeclareGraphicsExtensions{.eps}
\newcommand{\beq}{\begin{equation}}
\newcommand{\eeq}{\end{equation}}
\newcommand{\bea}{\begin{eqnarray}}
\newcommand{\eea}{\end{eqnarray}}
\newcommand{\ben}{\begin{eqnarray*}}
\newcommand{\een}{\end{eqnarray*}}
\newcommand{\bfig}{\begin{figure}}
\newcommand{\efig}{\end{figure}}

\usepackage{hyperref}
\hypersetup{
    colorlinks=true,      
    urlcolor=blue,
    citecolor=blue,
    linkcolor=blue
}

\definecolor{bpd}{rgb}{1,0,1}

\begin{document}
\title{Competing insulating phases of dipolar bosons in a dimerized optical lattice}
\author{Aoi Hayashi} 
\altaffiliation{Present address: School of Multidisciplinary Science, Department of Informatics, SOKENDAI (the Graduate University for Advanced Studies), 2-1-2 Hitotsubashi, Chiyoda-ku, Tokyo 101-8430, Japan ; National Institute of Informatics, 2-1-2 Hitotsubashi, Chiyoda-ku, Tokyo 101-8430, Japan}
\affiliation{Department of Physics, School of Science, Tokyo Institute of Technology, 2-1-2-1, Ookayama, Meguro-ku, Tokyo 152-8550, Japan}
\author{Suman Mondal}
\altaffiliation{Present address: Institut f\"{u}r Theoretische Physik, Georg-August-Universit\"{a}t G\"{o}ttingen, D-37077 G\"{o}ttingen, Germany}
\affiliation{Department of Physics, Indian Institute of Technology, Guwahati-781039, India}

\author{Tapan Mishra}
\email{mishratapan@iitg.ac.in}
\affiliation{Department of Physics, Indian Institute of Technology, Guwahati-781039, India}
\affiliation{Centre for Quantum Engineering Research and Education,
TCG Centres for Research and Education in Science and Technology, Sector V, Salt Lake, Kolkata 70091, India}

\author{B. P. Das}
\affiliation{Department of Physics, School of Science, Tokyo Institute of Technology, 2-1-2-1, Ookayama, Meguro-ku, Tokyo 152-8550, Japan}
\affiliation{Centre for Quantum Engineering Research and Education,
TCG Centres for Research and Education in Science and Technology, Sector V, Salt Lake, Kolkata 70091, India}

%

\date{\today}

\begin{abstract}
We study the ground state properties of dipolar bosons in a one dimensional dimerized optical lattice. In the limit of strong onsite repulsion i.e. hardcore bosons, and strong dipole-dipole interaction, a stable density wave (DW) phase is obtained at half filling as a function of lattice dimerization. Interestingly, at quarter filling we obtain the signatures of an insulating phase which has both the character the bond-order (BO) and the DW insulators which we call a bond-order density wave (BODW) phase. Moreover, we show that for a fixed hopping dimerization there occurs a BO-DW phase crossover as a function of the dipole-dipole interaction and the BODW phase is more robust when the hopping dimerization is stronger. We further examine the stability of the BODW phase in the limit of finite onsite interactions. 
\end{abstract}

\maketitle
\section{Introduction}
The systems of ultracold atoms in optical lattices have been the most versatile quantum simulator enabling to achive various novel phenomena of nature. Since the seminal observation of the superfluid (SF) to Mott insulator (MI) phase transition~\cite{Greiner_Bloch} of neutral bosonic atoms in optical lattices, enourmous progress has been made in the last two decades leading to the observation of a multitude of physical phenomena. The precise control over the system parameters acheived by suitable manipulation of lattice pontentials and/or the powerful technique of Feshbach resonance have made these systems capable of accessing limits which are difficult to achive in conventional solid state systems. This has led to a rapid surge in exploring physics both theoretically, most specifically in the context of Bose-Hubbard model and its variants and experimentally as well~\cite{TakahashiRev2020,BlochRev2017,BlochRev2012,lewensteinbook}

One of the most important variants of the BH model is the extended Bose-Hubbard (EBH) model which explains the physics of dipolar bosons, which can arise in polar atoms and molecules, Rydberg excited atoms in optical lattices with only nearest-neighbour (NN) interction. This simple model has been the topic of immense importance in the past~\cite{Baranov2008} and has attracted a renewed interest recently due to its experimental observation~\cite{Ferlaino2016} in optical lattices using Dysprosium (Dy) atoms. It has been well established already that the quantum phase diagram corresponding to the EBH model exhibits the gapped density wave (DW)  phases at commensurate densities characterized by the crystalline order. Interestingly, the supersolid phases (SS) appear at incommensurate densities which posses both the SF and the DW orders as a result of competing onsite and NN interactions~\cite{SantosReview2009,Baranov2008}. Apart from these two phases another interesting manifestation of this model is the Haldane insulator (HI) which exhibits a finite string order parameter~\cite{AltmanHI2006,AltmanHI2008,Fazio2012}.

On the other hand optical superlattices which are formed by superimposing two optical lattices of different wavelengths have shown to reveal a wealth of new physical phenomena recently~\cite{Porto1,Porto2,Porto3,Foelling, manpreetsuperlat1,manpreetsuperlat2,aryasuperlat1,aryasuperlat2,roth1,roth2,roux1,piel1,sebby1,cheinet1,shuchen1}. The modified periodicity of overall lattice in the process favours interesting ground state phenomena such as the onset of gapped phases at incommensurate densities associated to the primary lattice~\cite{manpreetsuperlat1,manpreetsuperlat2,roth1,roth2,danshita8,sayan1, shuchen1,Mondal2019}, symmetry protected topological phase transition~\cite{Grusdt2013,sumantopo1,sumantopo2,Ryu2002,DiLiberto2016}, frustrated magnetism~\cite{bloch_flux, Mishra2013_2}, disorder induced phase transition~\cite{DasSarmaBloch,BlochMBL2015,DasSarma2020,Roati2008,Ashirbad2021} and recently in the context of quantum computation~\cite{Jian-WeiPan2020}. One of the many variants of the superlattices is the double-well optical lattice which can be formed by superimposing a secondary lattice with wavelength twice that of the primary lattice. This particular superlattice also known as the dimerized lattice ensures that the hopping strengths of the particles alternate between the NN bonds. The physics of double-well lattice (hereafter called as dimerized lattice) is extremely important in the context of condensed matter physics~\cite{ssh}. 

For the case of noninteracting fermions and hardcore bosons, the  dimerized lattice manifests the interesting symmetry protected topological phase transition  which has been widely discussed in the framework of the celebrated Su-Schrieffer-Heeger (SSH) model~\cite{ssh}. Depending on the hopping dimerization, the system exhibits a gapped bulk spectrum and polarized zero energy edge modes characterized by the Zak phase~\cite{Zak1989,BlochZakPhase}. The bulk state exhibits finite oscillation in the bond kinetic energy and the system is known to be in the dimer or BO phase~\cite{Mishra2011,Mishra2013,Mishra2013_2,Mishra2014,MishraPolar2015}. Recently, the role of interaction has been investigated in the context the dimerized BH model~\cite{Grusdt2013,DiLiberto2016,DiLiberto2017,Mondal2019,sumantopo1,Hatsugai2020} predicting the bulk and edge properties of interacting bosons in optical lattices. A system of three-body constrained (TBC) boson (allowing a maximum of two bosons per site) on a one dimensional dimerized lattice was investigated by some of us in Ref.~\cite{Mondal2019}. It was shown that an attractive onsite interaction leads to a BO phase of bound bosonic pairs known as the pair-bond-order (PBO) phase at unit filling which crosses over to an MI phase in the limit of repulsive interaction. However, at half filling, only a gapped BO phase is stabilized in the repulsive regime. On the other hand, the effect of dimerized NN interaction has been studied extensively for systems of spin-polarized fermions or hardcore bosons~\cite{Kohmoto1981}. It has been shown that the system goes from the BO phase to a DW phase (phase separation) through an SF phase for repulsive (attractive) NN interaction at half filling. However, the investigation with uniform NN interaction and dimerized hopping at other densities and with finite onisite interaction  have not been explored in detail. 

\begin{figure}[t]
\centering
{\includegraphics[width=1\columnwidth]{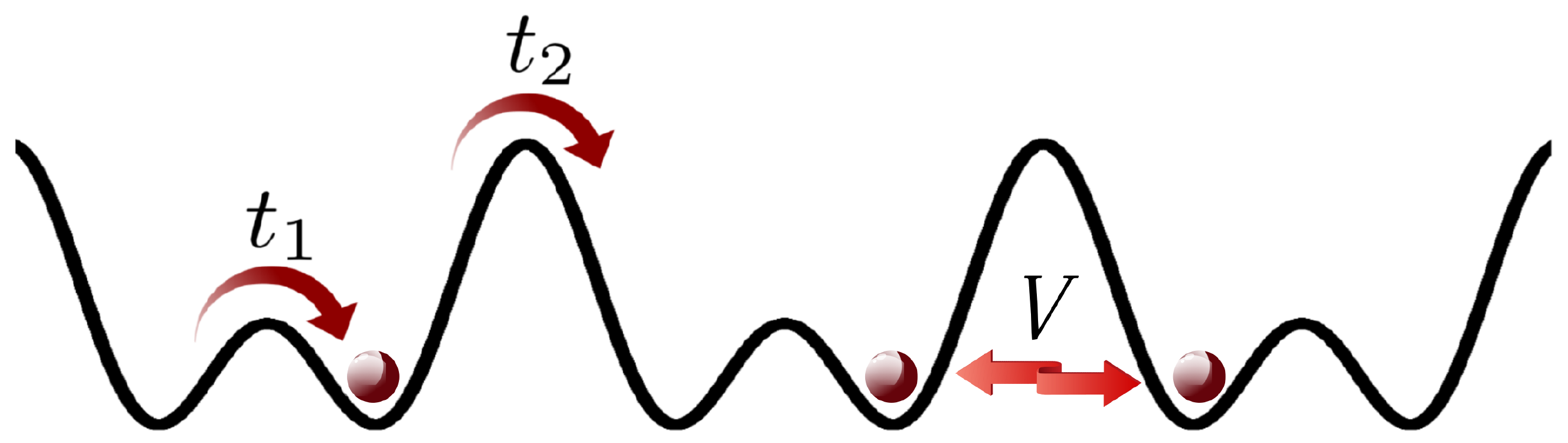}}
\caption{Double-well lattice structure with dimerized hopping $t_1 > t_2$. $V$ represents the NN interaction between the particles.}
\label{fig:lattice}
\end{figure}

In this paper we aim to fill the gap by studying the physics of hardcore dipolar bosons loaded onto a one dimensional dimerized optical lattice in the framework of the EBH model as depicted in Fig.~\ref{fig:lattice}. We analyse the interplay between the hopping dimerization and the NN interactions to explore the emergence of different insulating phases in the ground state of the system. Before going to the details of the results, we briefly highlight the important finding of our analysis. By considering a strong NN interaction, a change in the dimerization results in an insulating phase at quarter filling which exhibits both the BO and DW orders. This insulating phase is found to be very sensitive to the change in dimerization strength for fixed NN interactions.  Apart from this we have obtained the signatures of the BO and DW phases at half filling and transitions between them. In the end we examine the stability of the non-trivial insulating phase at quarter filling in the limit of finite onsite interactions.

The remaining part of the paper is organized as follows. In Sec~\ref{sec:model} we discuss the model which is considered in our studies and the method adopted. We present the results and discussion in Sec.~\ref{sec:res} follwed by the conclusions in Sec.~\ref{sec:con}. 

\section{Model and approach}\label{sec:model}

The EBH model for a dimerized lattice bosons is given by 
\begin{eqnarray}
H=&-& t_1 \sum_{i\in odd} \left(a_{i}^{\dagger}a_{i+1}^{\phantom \dagger} + \text{H.c.}\right)- t_2\sum_{i\in even}\left(a_{i}^{\dagger}a_{i+1}^{\phantom \dagger}+\text{H.c.}\right)\nonumber\\ 
&+& \frac{U}{2}\sum_i n_i(n_i-1)+V \sum_{i}  n_{i} n_{i+1}
\label{eq:ham}
\end{eqnarray}
where, $a_i$ and $n_i$ are the bosonic annihilation and number operator respectively at sites $i$. $t_1$ and $t_2$ are intra- and inter-cell hopping strengths. $U$ and $V$ are the on-site and nearest-neighbor interaction energies. The hopping dimerization is introduced by defining $\delta=t_2/t_1$ and setting $t_1 >  t_2$. In the entire simulation we set $t_1=1$ which makes all the physical quantities dimensionless.

In our studies, we impose hardcore constraint on bosons by assuming $a_i^{\dagger 2} = 0$ which can be achieved in the limit of $U\to\infty$.  . The physics of the model shown in Eq.~\ref{eq:ham} in some limiting situations are well known at half filling. 
After a Jordan-Wigner transformation to free fermions, the Eq.~(\ref{eq:ham}) maps to the interacting SSH model. As highlighted before, in the limit $V=0$ (the SSH model), the model~(\ref{eq:ham}) exhibits a BO phase for any $\delta \neq 1$  at half-filling~\cite{Mondal2019}. On the other hand, it is well known that in the absence of any dimerization (i.e. $\delta=1$), the model~(\ref{eq:ham}) can be suitably maped to the XXZ model that exhibits a gapless SF to insulating DW phase transition at a critical NN interaction of $V=2t$ at half filling. In this work, our aim to explore the emergence of the insulating phases and their nature that may result from the interplay of both $\delta$ and $V$. 

To explore the ground state properties of the many-body  Hamiltonian given in Eq.~\ref{eq:ham} we employ the Matrix Product State (MPS) based Density Matrix Renormalization Group (DMRG) method with a maximum bond dimension of $D = 500$. We consider system sizes up to $L=240$ and explore the physics at different densities $\rho=N/L$ of interest where $N$ is the total number of bosons in the system. We calculate all the physical quantities in the thermodynamic limit ($L=\infty$) using appropriate finite size extrapolation, unless otherwise mentioned. 

\section{Results}\label{sec:res}
This section is devided into three parts. To understand to role of interaction, in the first part, we consider a fixed interaction $V$ and explore the existence of the insulating phases as a function of $\delta$ at different densities. In the second part, we explore the effect of $V$ by fixing $\delta$. In the end we study the effect of finite onsite interaction for some exemplary values of $\delta$ and $V$.  
\begin{figure}[t]
\begin{center}
\includegraphics[width=\linewidth]{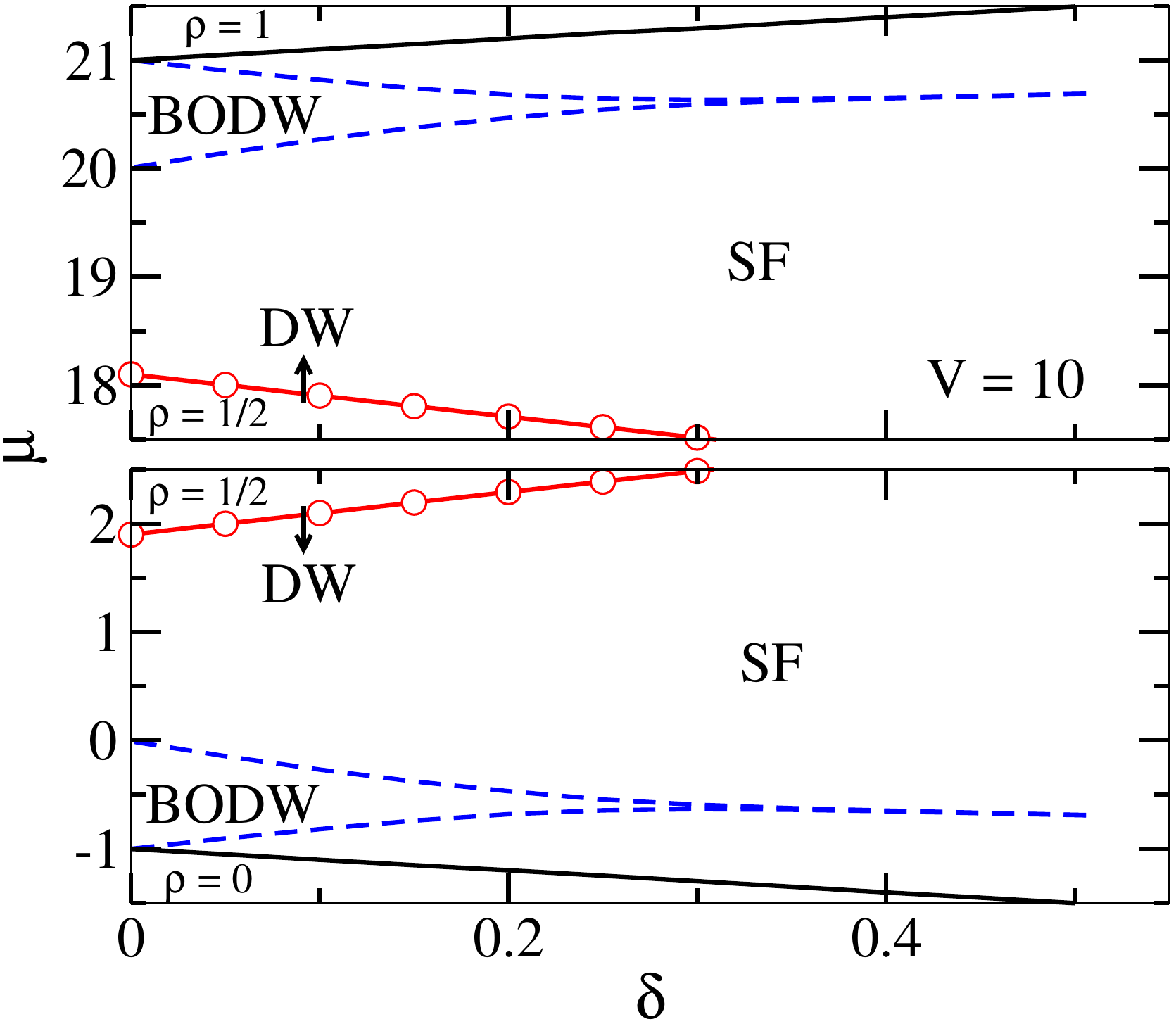}
 \end{center}
\caption{The phase diagram depicting the insulating phases in the $\mu-\delta$ plane ($t_1 = 1$) for the hardcore bosons with $V=10$. The gapped phases at $\rho = 1/4$ and $3/4$ are the BODW phases bounded by the black lines with circles, whereas at $\rho=1/2$, the phase is the DW phase, bounded by red lines with squares. The empty ($\rho = 0$) and full ($\rho = 1$) states are separated by the blue lines with triangles. }
\label{fig:pdHC}
\end{figure}

\subsection{Fixed Interaction}\label{sec:int}

In this subsection, we study the combined effect of both hopping dimerization (i.e $\delta\neq1$) and NN interaction (i.e. $V\neq0$) to explore the possibilities of insulating phases in the system without restricting ourselves to half filling. For this purpose, we consider  a very strong NN interaction i.e. $V=10$ and obtain the ground state phase diagram of model~(Eq.~\ref{eq:ham}) which is shown in Fig.~\ref{fig:pdHC}. Interestingly, the phase diagram of Fig.~\ref{fig:pdHC} shows three gapped phases corresponding to $\rho=1/4,~1/2$ and $3/4$. The gapped phases are extracted from the behavior of the single particle excitation gap 
\begin{equation}
 \Delta_L = \mu^+_L-\mu^-_L
\end{equation}
where $\mu^+_L=E_L(N+1)-E_L(N) ~\rm{and}~ \mu^-_L=E_L(N)-E_L(N-1)$ are the chemical potentials. Here, $E(N)$ denotes the ground state energy of the system with $N$ particles. It can be seen from Fig.~\ref{fig:pdHC} that the gap remains finite at $\rho=1/2$ for all values of $\delta$ considered (region bounded by the red circles). However, for $\rho=1/4 $ and $3/4$ there is a phase transition to a gapless region as indicated by the smooth closing of the gap $\Delta_L$ as a function of $\delta$ (region bounded by the blue dashed curves). The solid black lines at the top and bottom in the phase diagram of Fig.~\ref{fig:pdHC} correspond to the empty and full states. 
\begin{figure}[t]
\begin{center}
\includegraphics[width=1\columnwidth]{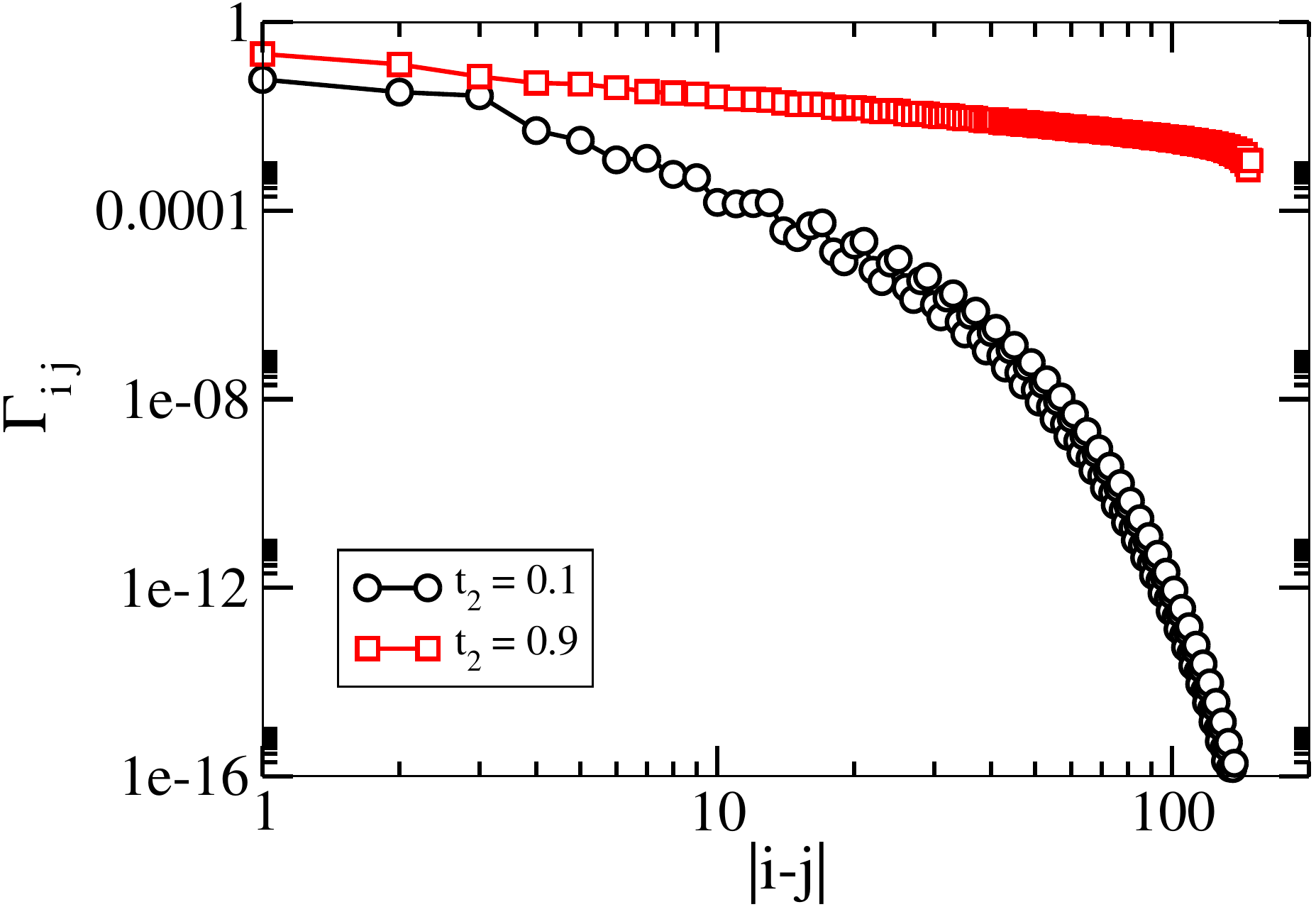}
 \end{center}
\caption{The correlation function $\Gamma_{ij}$ is plotted with distance $|i-j|$ in log-log scale for strong dimerization ($\delta = 0.1$) and weak dimerization ($\delta = 0.9$) at $\rho = 1/4$ of a finite system of size $L=240$. The exponential decay of the curve ($\delta = 0.1$) signifies the gapped phase and a power law decay ($\delta = 0.9$) indicates a  gapless phase.}
\label{fig:corr}
\end{figure}
The gapped and gapless phases can be confirmed by computing the single particle correlation function which is given by
\begin{equation}
\Gamma_{ij} = \langle a_i^\dagger a_j\rangle. 
\end{equation} 
In Fig.~\ref{fig:corr} we plot $\Gamma_{ij}$ against the distance $|i-j|$ in the log-log scale for $\delta=0.1$ and $0.9$ which are within the gapped and gapless regions respectively at $\rho=1/4$ in the phase diagram of Fig.~\ref{fig:pdHC}. 
The exponential (powerlaw) decay of $\Gamma_{ij}$ for $\delta=0.1 (0.9)$ indicates the gapped (gapless) phase. 

\begin{figure}[t]
\begin{center}
\includegraphics[width=1\columnwidth]{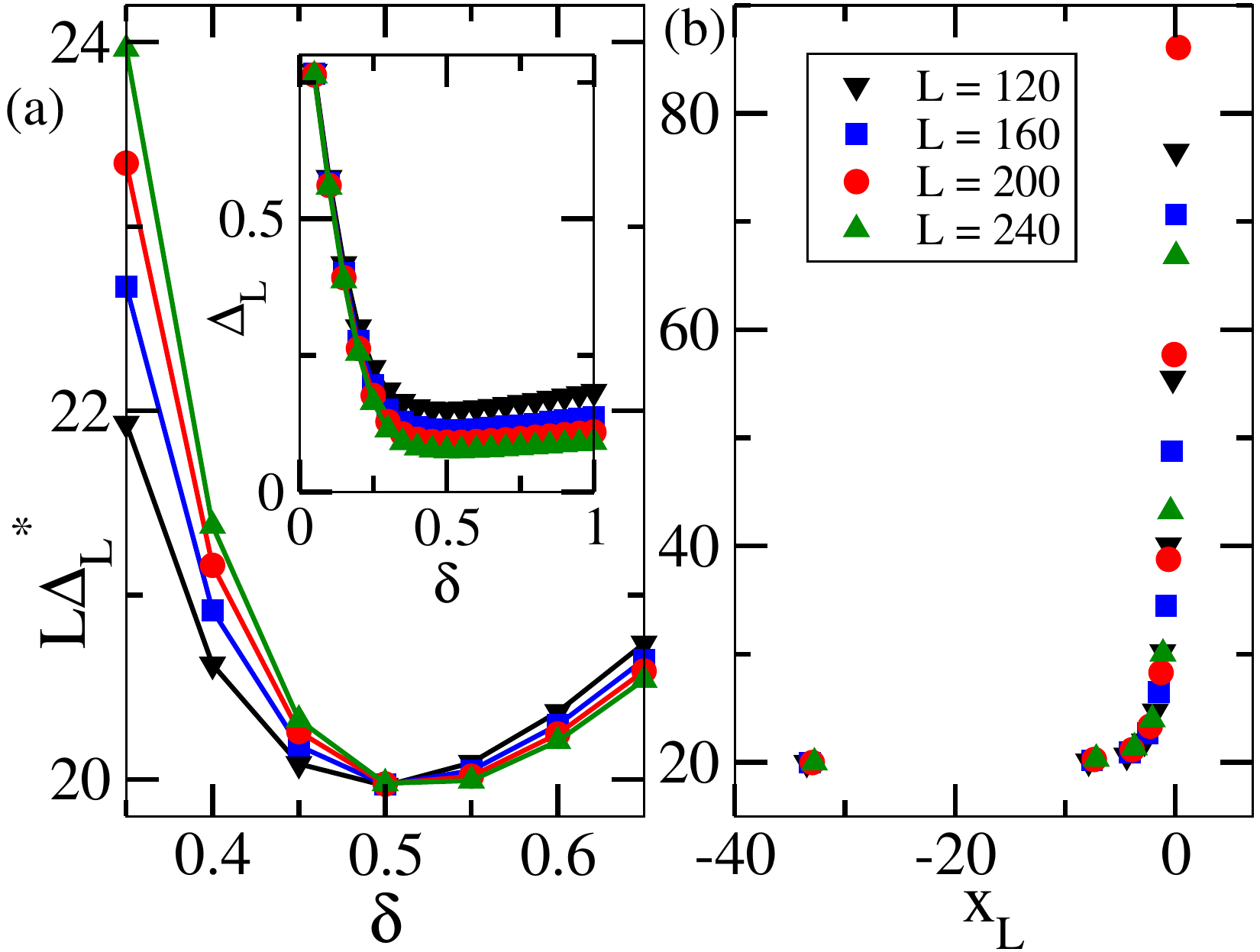}
 \end{center}
\caption{The finite-size scaling of $\Delta_L$ is shown to find the BKT transition point. (a) A perfect crossing of all the curves for different $L$ at $\delta \sim 0.506$ represents the critical point. (b) The collapse of all the data for different $L$ near the critical point ($x_L = \infty$) confirms the BKT transition with $\delta_{c} \sim 0.506$.  The inset shows the $\Delta_L$ as a function of $\delta$ for different $L$ indicating the gap minimum at $\delta \sim 0.5$.}
\label{fig:gap_scale}
\end{figure}

Although $\Gamma_{ij}$ provides insights about the nature of the phases, it is hard to obtain the phase transition critical points in a situation where the gap or the correlation function varies rather smoothly. This kind of signatures at $\rho=1/4$ and $3/4$ in this case is typical for one dimensional systems which indicate a Berezinskii-Kosterlitz-Thouless (BKT) type transition~\cite{RigolRev2011}. To accurately quantify the transition points we utilize the finite-size scaling of the gap $\Delta_L$ to find the critical $\delta$ of transition following Ref.~\cite{tvv}. According to the scaling theory, the quantity 
\begin{equation}
L\Delta_L^* = L\Delta_L \left( 1+ \frac{1}{2\rm{ln}L +C} \right)
\end{equation}
must be length invariant at a critical value of $\delta$ and at the same time exibits a perfect data collapse as a function of $x_L= \rm{ln} L - \frac{a}{\sqrt{|\delta -\delta_{c}|}}$ near the critical point ($\delta_{c}$) for suitable values of $C$ and $a$.
As shown in Fig.~\ref{fig:gap_scale}, for $\rho=1/4$, the perfect crossing of all the curves (Fig.~\ref{fig:gap_scale}(a)) at $\delta \sim 0.506$ and collapse of all the data for different $L$ near the critical point (Fig.~\ref{fig:gap_scale}(b)) indicate a BKT type phase transition to the gapped phase at a critical point $\delta_c \sim 0.506$. 

\begin{figure}[t]
\begin{center}
\includegraphics[width=1\columnwidth]{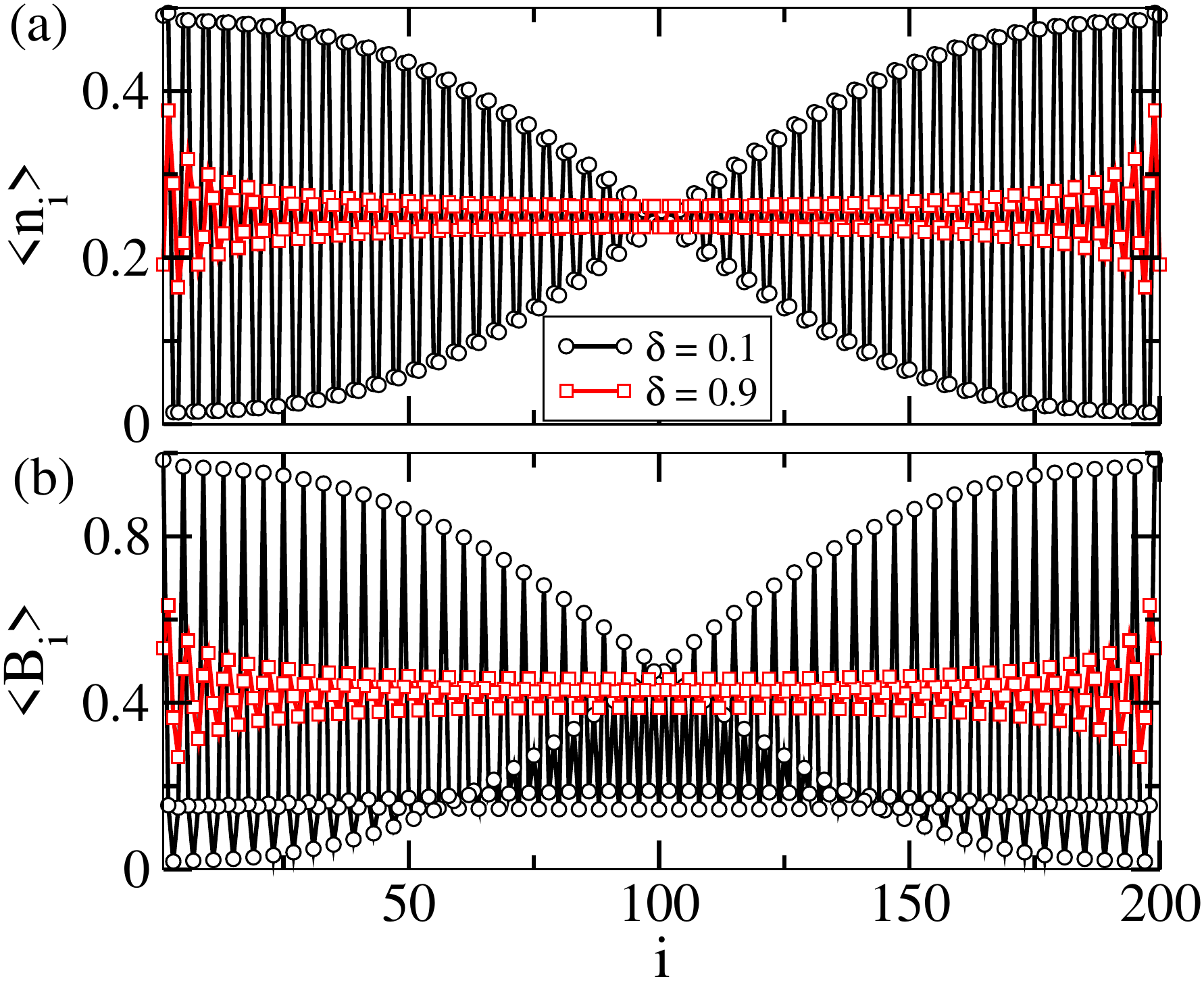}
 \end{center}
\caption{The values of $\langle n_i\rangle$ and $\langle B_i\rangle$ are plotted in (a) and (b) respectively as a function of the site index $i$ at $\rho = 1/4$ for a system of length $L=200$ and $V=10$. The finite oscillations in both  $\langle n_i\rangle$ and $\langle B_i\rangle$ for $\delta = 0.1$ (black circles) represent the simultaneous existence of both DW and BO orders respectively. The oscillations die out for $\delta = 0.9$ (red squares) which indicate the vanishing of the DW and BO orders.}
\label{fig:ni_boi}
\end{figure}

After separating the insulating phases at different densities, we now focus on to identify the nature of these phases. We find that the gapped phase at $\rho=1/2$ is a DW phase characterized by a finite oscillation in the real-space density $<n_i>$. The DW nature can be well understood by using the density structure factor which is given by 
\begin{equation}
S(k) = \frac{1}{L^2}\sum_{i,j}e^{ikr}(\langle n_{i}n_{j}\rangle - \langle n_{i}\rangle \langle n_{j}\rangle)
\label{eq:strd}
\end{equation}
where $r=|i-j|$ and $k$ is the crystal momentum. In our simulation we find finite peaks at $k=\pm\pi$ in the structure factor for all values of $\delta$ considered, indicating a DW phase where the wavefunction is a product state with one particle in every other sites i.e.
\begin{equation}
|\psi\rangle_{DW} = |.~.~.~1~0~1~0~1~0~1~.~.~.\rangle.
\end{equation}
However, we notice that the insulating phases at $\rho=1/4$ and $3/4$ are of different nature. Interestingly, they exhibit both BO and DW order in their charachter. 
\begin{figure}[b]
\begin{center}
\includegraphics[width=1\columnwidth]{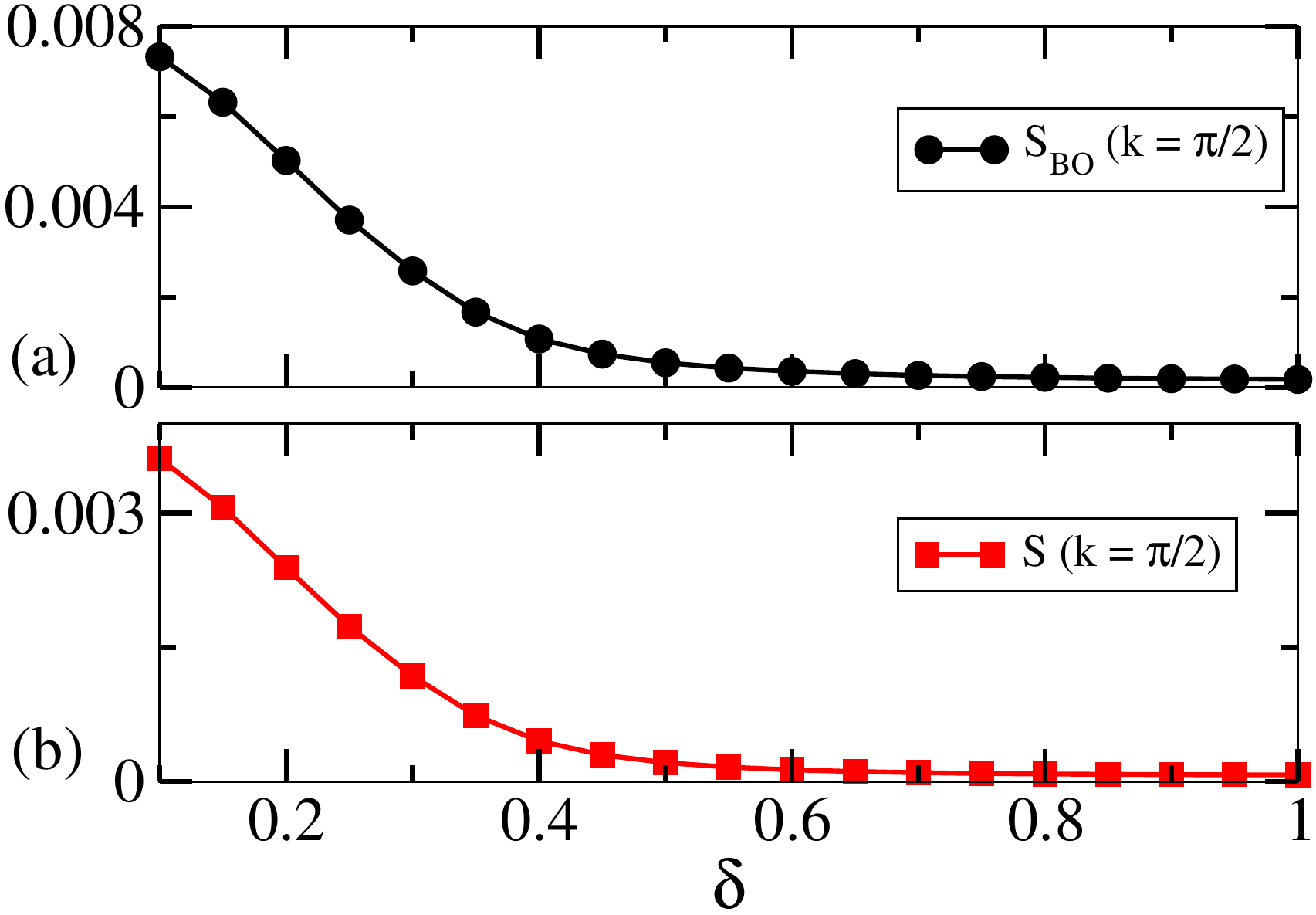}
 \end{center}
\caption{The extrapolated values of $S_{BO}(k=\pi/2)$ and $S(k=\pi/2)$ are plotted in (a) and (b) respectively with $\delta$ at $\rho = 1/4$.}
\label{fig:ops_hc1}
\end{figure}
To characterise this we calculate the expectation values of $n_i$ and the bond energy 
$B_i = a_i^\dagger a_{i+1} + \rm H.c.$ which are plotted in Fig.~\ref{fig:ni_boi}(a) and (b) respectively for $\delta = 0.1$ and $0.9$. The strong oscillations for $\delta=0.1$ in both $\langle n_i\rangle$ and $\langle B_i\rangle$ indicate the signatures of both BO and DW orders respectively. In contrast, the oscillations tend to die out for $\delta=0.9$ although they remain finite due to finite size effect. The signatures of DW and BO orders can be well understood from the finite peak in the density structure factor $S(k)$ (Eq.~\ref{eq:strd}) as well as the BO structure factor $S_{BO}(k)$ which is defined as 
\begin{equation}
S_{BO}(k) = \frac{1}{L^2}\sum_{i,j}e^{ikr}\langle B_{i}B_{j}\rangle 
\label{eq:strb}
\end{equation}
at finite values of $k$. For $\rho=1/4$, we obtain a sharp peak at $k=\pi/2$ for all values of $\delta$ within the gapped region of Fig.~\ref{fig:pdHC}. This behavior suggests that within the gapped region at $\rho=1/4$, the particles are located at every alternate double-wells in the lattice due to strong repulsive $V$ and within each occupied double-well there is a finite bond energy $B_i$. We call this insulating phase the bond-order-density wave phase or BODW phase. As the value of $\delta$ becomes larger or the dimerization becomes weaker, the bosons can no more be trapped in the double-well. In this limit the bosons can freely move throughout the lattice owing to their incommensurate density turning the system a gapless SF phase. The BODW to SF phase transition can further be quantified by looking at the behavior of the BO and DW structure factors for different values of $\delta$.
In Fig.~\ref{fig:ops_hc1}(a) and (b) we plot the finite size extrapolated values of peak heights of $S_{BO}(k=\pi/2)$ and $S(k=\pi/2)$ as a function of $\delta$. The extrapolation is done by using systems of different sizes where the maximum system size is $L = 240$. The vanishing of both $S_{BO}(k=\pi/2)$ and $S(k=\pi/2)$ roughly at $\delta \sim 0.5$ indicates a BODW-SF transition that matches well with the critical point obtained using the gap scaling (see Fig.~\ref{fig:gap_scale}). 
Note that similar to the $\rho=1/4$ case, the gapped phase at $\rho=3/4$ is also found to be a BODW phase.

\begin{figure}[b]
\begin{center}
\includegraphics[width=1\columnwidth]{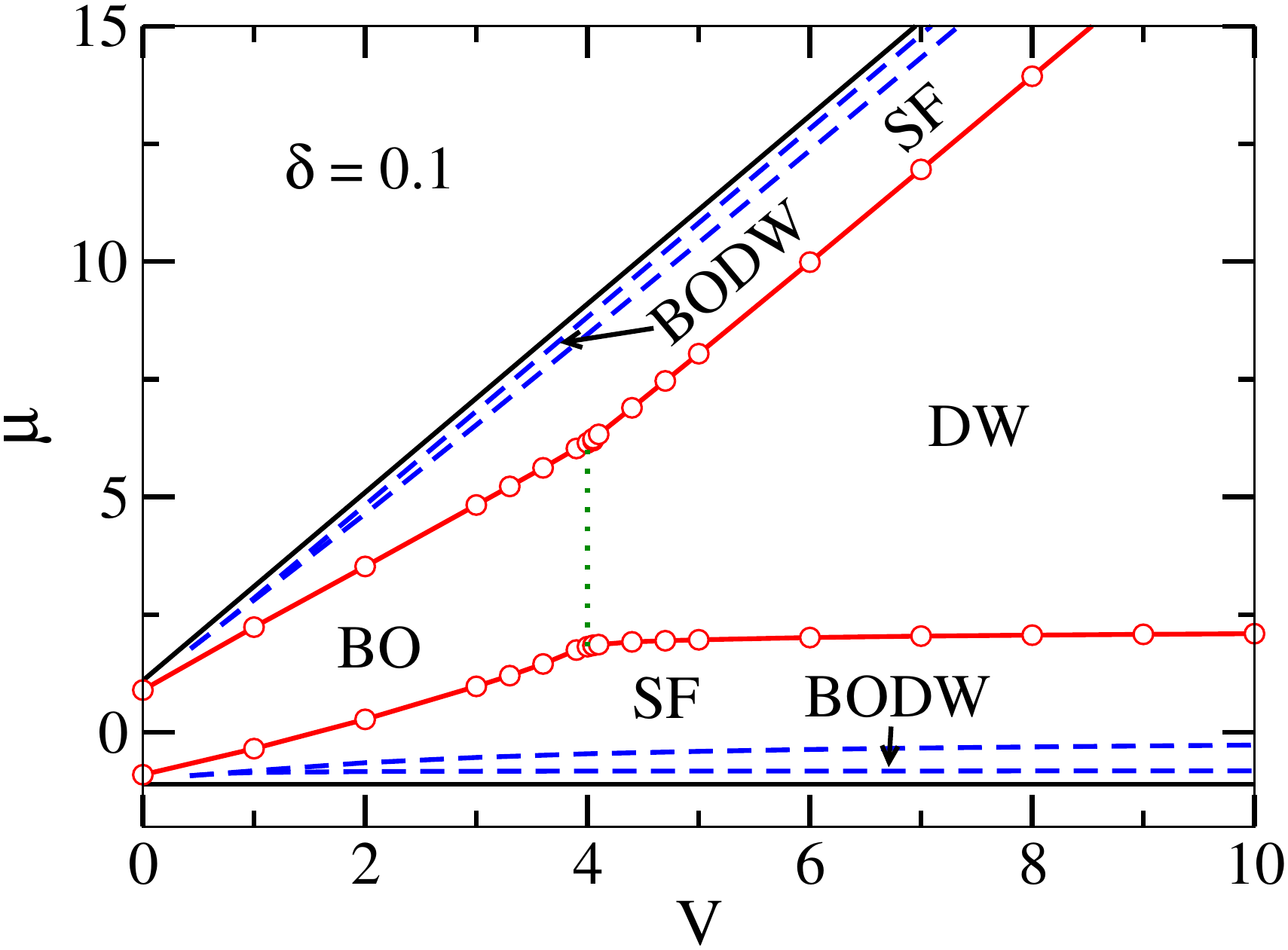}
\end{center}
\caption{The phase diagram showing the insulating phases for the hardcore bosons in the $\mu-V$ plane ($t_1 = 1$) for $\delta = 0.1$. The insulating phases at $\rho = 1/4$ and $3/4$ are the  BODW phases (bounded by the dashed blue lines). At $\rho=1/2$, the BO-DW transition is marked by the dotted line. }
\label{fig:pdHCV}
\end{figure}

\subsection{Fixed Dimerization}\label{sec:int}
\begin{figure}[t]
\begin{center}
\includegraphics[width=1\columnwidth]{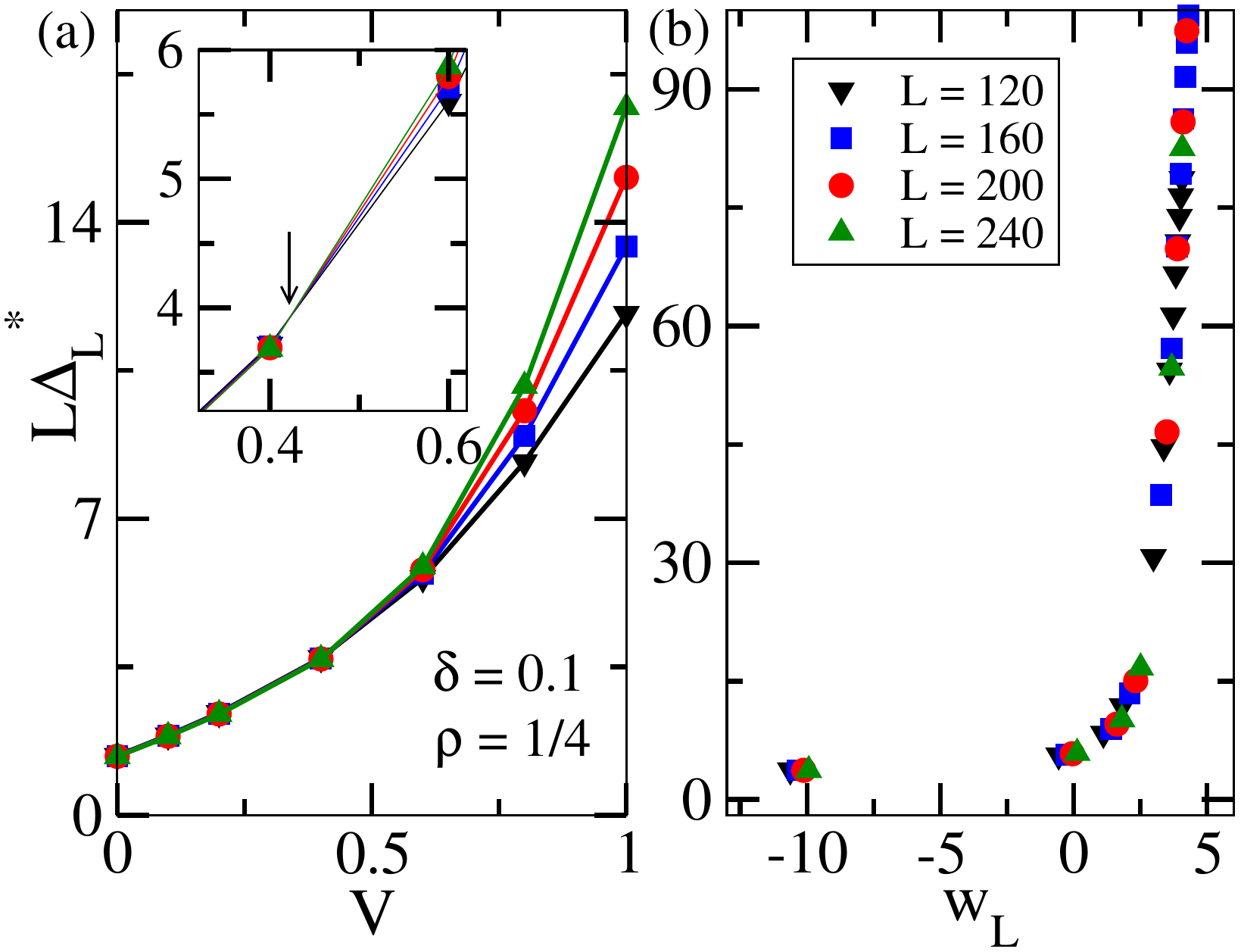}
\end{center}
\caption{The finite-size scaling of $\Delta_L$ is shown to find the BKT transition point. (a) A perfect crossing of all the curves for different $L$ at $V \sim 0.42$ represents the critical point. The inset shows the magnified plot near the critical point. (b) The collapse of all the data for different $L$ near the critical point ($w_L = \infty$) confirms the BKT transition with $V_{c} \sim 0.42$.}
\label{fig:gap_scale2}
\end{figure}
After obtaining the behaviour of the system by varying $\delta$ for a fixed large $V$, we explore the physics for a fixed $\delta$ and varying $V$. In this regard, we consider a case of strong dimerization such as $\delta = 0.1$, where the half-filled and quarter-filled sectors belong to the gapped phases (see Fig.~\ref{fig:pdHC})  and vary $V$. As already discussed above,  in the absence of $V$ and $\delta=0.1$, there exists only an insulating BO phase at half filling. On the other hand, for large $V$, at quarter- (half-) filling, the system is in the BODW (DW) phase. Our analysis reveals how these  two limits interpolate as a function of $V$ resulting in a phase diagram as shown in  Fig.~\ref{fig:pdHCV} in the $V-\mu$ plane. As the value of $V$ increases, two BODW phases start to appear (opening of the gap) at $\rho=1/4$ and $3/4$ and the lobes become bigger as a function of $V$.  
The phase transitions from the SF to the BODW phase at $\rho = 1/4$ and $3/4$ are found to be of BKT type. The critical points are calculated from the finite-size scaling of the $\Delta_L$ following the method discussed before. As the gap variation here is with respect to $V$, we replace $x_L$ by $w_L$ as,
\begin{equation}
w_L = \rm{ln} L - \frac{a}{\sqrt{|V -V_{c}|}}.
\end{equation}
In Fig.~\ref{fig:gap_scale2} we portray the gap scaling for $\rho = 1/4$ and $\delta = 0.1$. A crossing of all the curves at $V \sim 0.42$ in  Fig.~\ref{fig:gap_scale2}(a)  and a complete collapse of data for different $L$ in Fig.~\ref{fig:gap_scale2}(b) yields the BKT transition point at $V_c \sim 0.42$. Similar analysis for $\rho = 3/4$ with $\delta = 0.1$ shows the critical transition point at $V_c \sim 0.43$ as indicated in the phase diagram of Fig.~\ref{fig:pdHCV}. 

\begin{figure}[t]
\begin{center}
\includegraphics[width=1\columnwidth]{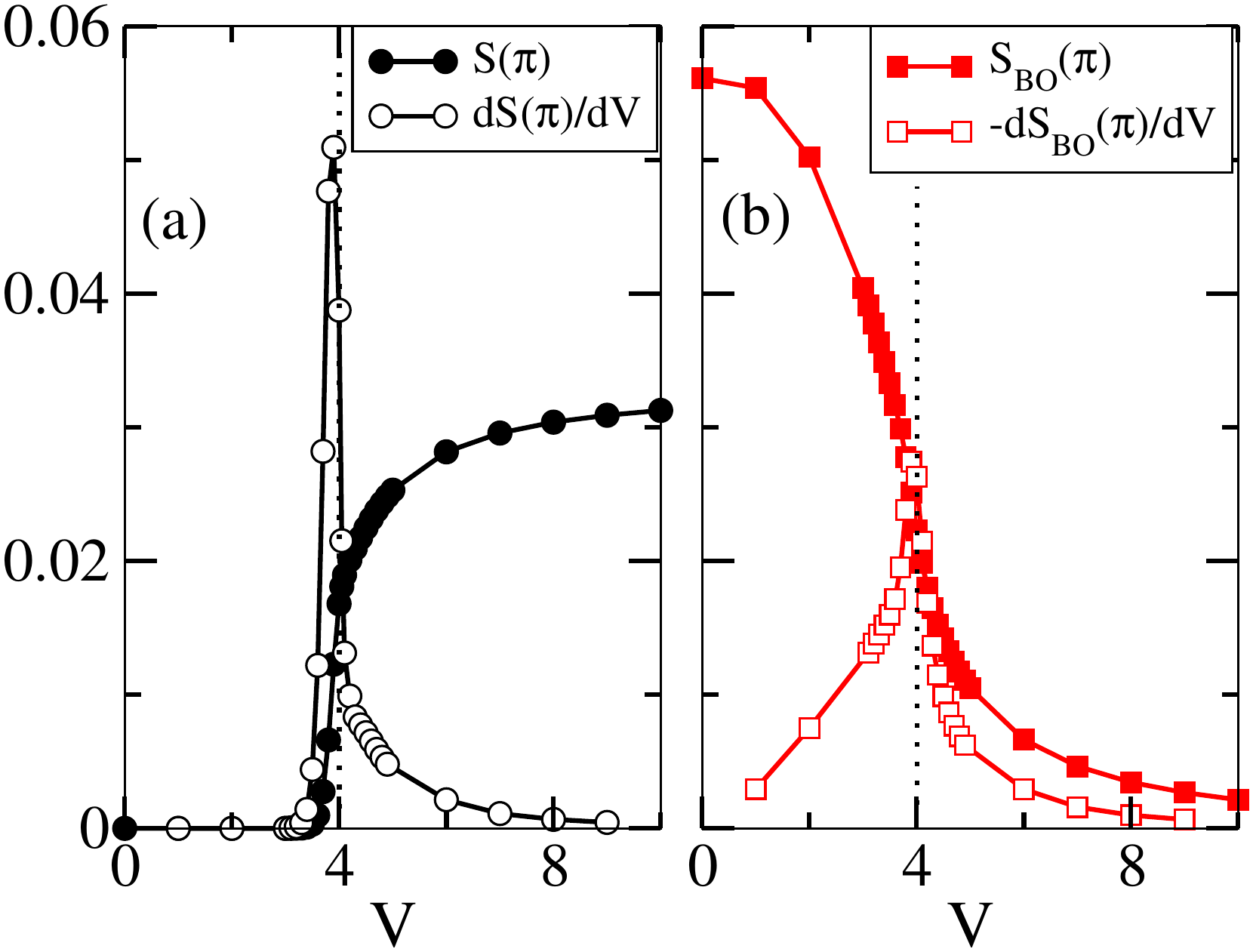}
\end{center}
\caption{(a) $S(\pi)$ with its derivative and (b) $S_{BO}(\pi)$ with its derivative are plotted with respect to $V$ for $\delta = 0.1$ and $\rho = 1/2$ corresponding to the Fig.~\ref{fig:pdHCV}. The vertical dotted lines at $V\sim 4.0$ correspond to the dotted line shown in Fig.~\ref{fig:pdHCV} which matches fairly well with the peaks of the derivative functions.}
\label{fig:ops_hc2}
\end{figure}
\begin{figure}[b]
\begin{center}
\includegraphics[width=1\columnwidth]{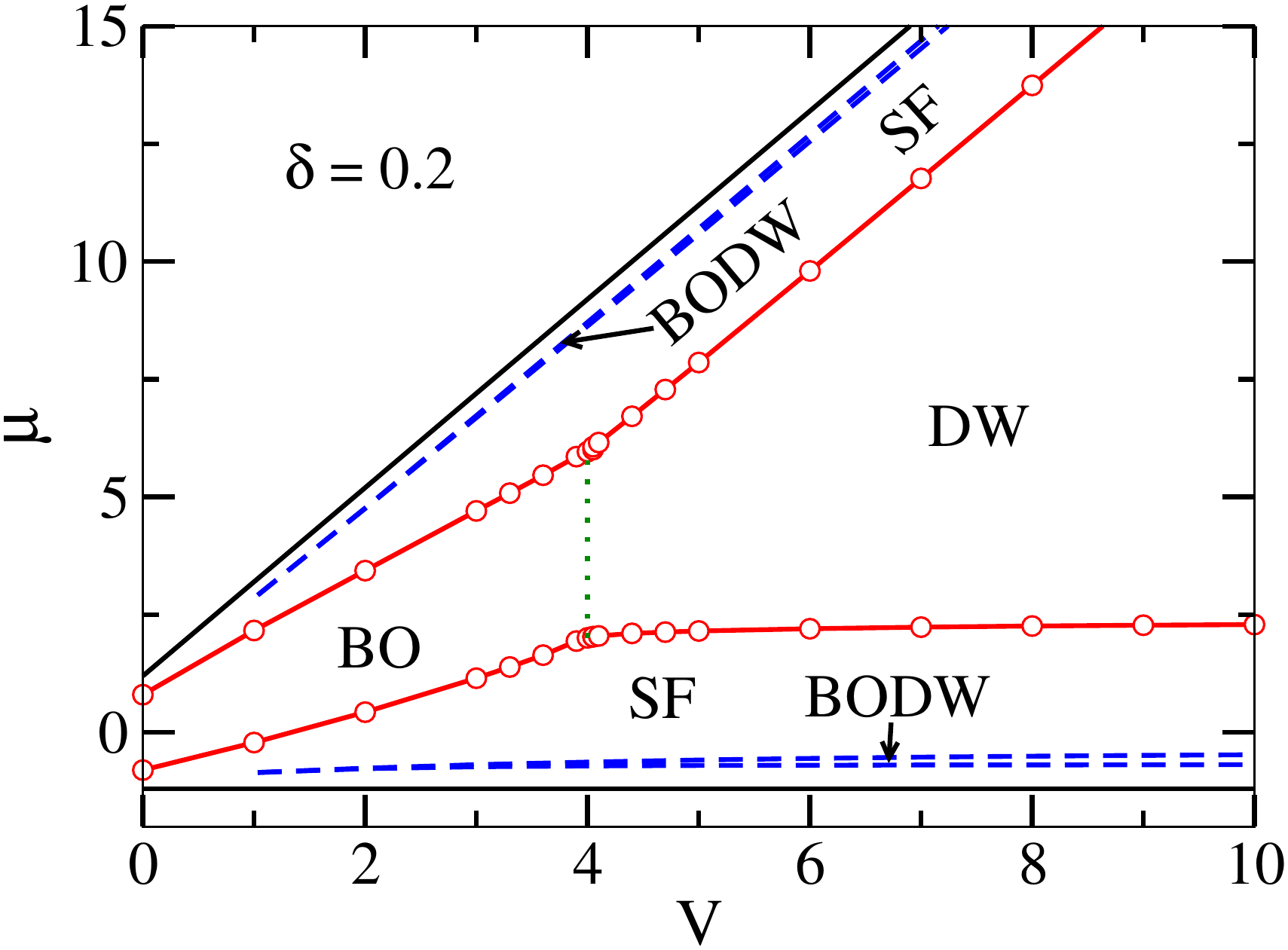}
\end{center}
\caption{The phase diagram showing the insulating phases of the hardcore bosons in the $\mu-V$ plane ($t_1 = 1$) for $\delta=0.2$. At $\rho = 1/4$ and $3/4$ the insulating phases are the BODW phases (the region bounded by the dashed blue lines). At $\rho=1/2$, there exists a BO-DW phase transition marked by the dotted line.}
\label{fig:pdHCV2}
\end{figure}
On the other hand, the BO phase at $\rho=1/2$ becomes a DW phase with increasing $V$ as the NN interaction dominates over the dimerization strength. As both the BO and DW phases are insulating phases, the BO-DW transition can be inferred from a kink in the gap $\Delta$ as a function of $V$. This feature is reflected as the kinks in the $\mu^+$ and $\mu^-$ curves (red circles in Fig.~\ref{fig:pdHCV}) at $V\sim 4$. The BO-DW transition point can be quantified by analysing the corresponding structure factors $S(k)$ and $S_{BO}(k)$. In Fig.~\ref{fig:ops_hc2}(a) and (b) we plot the extrapolated values  of $S(\pi)$ and $S_{BO}(\pi)$ as a function of $V$ for $\delta =0.1$. The vanishing $S(\pi)$ (filled circles) and finite $S_{BO}(\pi)$ (filled squares) in the regime of small $V$ in Fig.~\ref{fig:ops_hc2}(a)  and (b) respectively clearly indicate the signature of the BO phase. However, after a critical $V_c$, the $S(\pi)$ becomes finite which corresponds to the transition to the DW phase. In order to obtain the critical point for this BO-DW transition we use the derivative of $S(\pi)$ with respect to $V$. The plot of $dS(\pi)/dV$ as a function of $V$ exhibits a sharp peak at $V\sim 4$ as shown in Fig.~\ref{fig:ops_hc2}(a) as empty circles. Similar feature in the derivative of $S_{BO}(\pi)$ (empty squares) in Fig.~\ref{fig:ops_hc2}(b) confirms the BO-DW transition which is indicated as a dotted line in Fig.~\ref{fig:pdHCV}.


From the above analysis it is evident that the BODW phases exhibit smaller gaps compared to those of  the DW and BO phases. We find that the BODW lobe is extremely sensitive to dimerization $\delta$. By considering a slightly weaker dimerization i.e. $\delta=0.2$, we obtain the phase diagram as shown in Fig.~\ref{fig:pdHCV2}. It can be seen that the gaps at $\rho=1/4$ and $3/4$ are reduced, thereby shrinking the BODW lobes. However, the effects on the BO and DW phases at $\rho=1/2$ are very small.


\begin{figure}[b]
\begin{center}
\includegraphics[width=1\columnwidth]{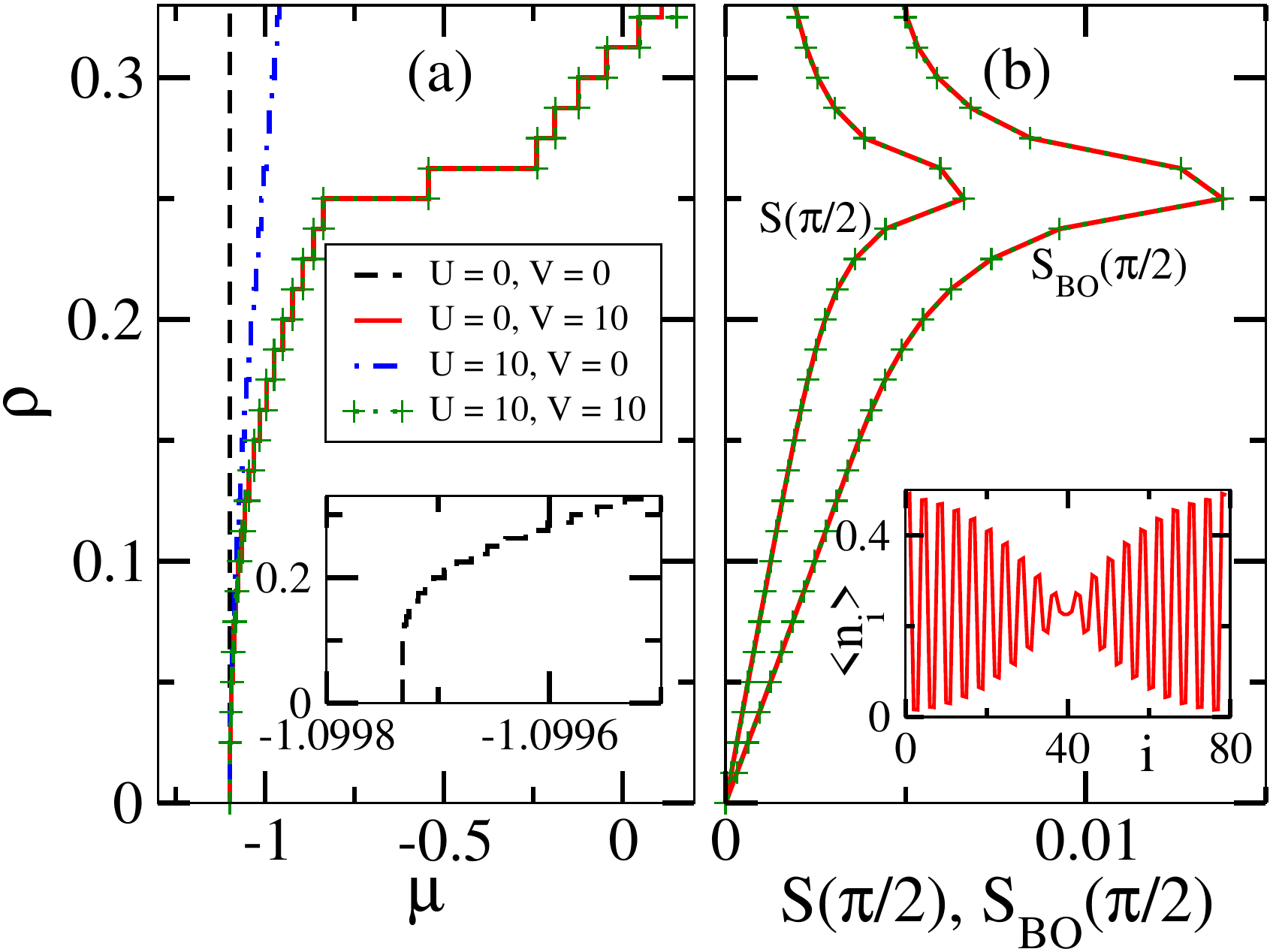}
\end{center}
\caption{(a)The variation of $\rho$ with increasing $\mu$ is plotted for different combinations of $U$ and $V$ using a finite system of size $L=80$ and $\delta=0.1$. The plateaus at $\rho = 1/4$ appear when $V=10$ indicating the gapped BODW phase. (b) The plots of $S(\pi/2)$ and $S_{BO}(\pi/2)$ as a function of $\rho$ for $U=0,~V=10$ and $U=10,~V=10$ confirm the stability of the BODW phase. The inset of (a) shows the zoomed in $\rho-\mu$ plot of the main figure for $U=0$ and $V=0$  for clarity. The inset of (b) shows the particle number distribution along the lattice  for $\rho =1/4$ with $V=10$ and $U=0$. }
\label{fig:rho_mu}
\end{figure}

\subsection{Finite onsite interaction}\label{sec:TBCSC}
Finally, in this section, we examine the stability of the BODW phase by relaxing the  the hardcore constraint. For this we allow finite onsite interaction $U$ of bosons in Eq.~\ref{eq:ham}. 
Here, we only discuss the fate of the BODW phase at $\rho = 1/4$. To this end we fix $\delta=0.1$ and calculate the $\rho-\mu$ curves for different combinations of $U$ and $V$ as shown in Fig.~\ref{fig:rho_mu}(a). The finite plateaus at $\rho=1/4$ for $V=10$ in the limit of small and large $U$ such as $U=0$ (red solid curve) and $U=10$ (green plus) clearly indicate an insulating phase. In order to confirm the nature of this phase, we plot the structure factors $S(\pi/2)$ and $S_{BO}(\pi/2)$ as a function of $\rho$ in Fig.~\ref{fig:rho_mu}(b) for both the combinations such as ($U=0,~V=10$) and ($U=10,~V=10$). The finite peaks at $\rho=1/4$ in both the structure factors confirms the stability of the BODW phase. In the inset of Fig.~\ref{fig:rho_mu}(b), we plot $\langle n_i\rangle$ at $\rho=1/4$ for $V=10$ and $U=0$ where each double-well has one particle followed by an empty double well, confirming the existence of the BODW phase. On the other hand, the $\rho~-\mu$ curves for $V=0$ do not show any plateaus which indicate that the system is in the gapless SF phases (see Fig.~\ref{fig:rho_mu}(a)). Note that this BODW phase appears due to the interplay between the long-range interaction and hopping dimerization alone and the role of $U$ is not significant.

\section{Conclusions}\label{sec:con}
In conclusion, we have studied the physics of the ground state of a system of interacting hardcore bosons in a one dimensional optical lattice with hopping dimerization. By using the DMRG method, we have predicted the emergence of various insulating phases and associated phase transitions. We have shown that in presence of large NN interaction, the system exhibits a DW phase which remains stable for all values of dimerization strengths. Interestingly, at quarter fillings, we have obtained the signatures of an insulating phase which exhibits both the bond order and density wave orders which we call the BODW phase. Moreover, we have obtained a SF-BODW phase transition as a function of the dimerization strength which is of the BKT universality class. On the other hand, by fixing the dimerization strength, there occurs a BO-DW phase transition at half filling as a function of the NN interaction. In this case also, we have obtained a SF-BODW phase transition at quarter filling. It is found that the BODW phase is extremely sensitive to the dimerization strength. In the end we have predicted the stability of the newly found BODW phase in the limit of finite onsite interaction. 

Our findings predict a new phase at quarter filling which is a result of both hopping dimerization and the NN interaction. As mentioned before, the lattice model we have considered resembles a double well optical lattice which has been realized in experiments in the context of topological phase transitions. Hence, our prediction can in principle be realized in quantum gas experiments~\cite{BlochZakPhase}. As a possible future direction, it will be interesting to explore the topological properties of the model considered at quarter filling. As the system exhibits a gap in this limit, it may favour a topological phase transition which can be studied by changing the hopping dimerization in the framework of the interacting SSH model~\cite{sumantopo1}.


\bibliography{references}

\end{document}